\documentclass[traditabstract,longauth,letter]{aa} 
\usepackage{graphicx}
\usepackage{txfonts}
\usepackage{natbib}
\usepackage{colordvi}
\newcommand{\mic}{~$\mu$m}
\newcommand{\micJy}{~$\mu$Jy}
\newcommand{\spitzer}{{\it Spitzer}}
\newcommand{\herschel}{{\it Herschel}}

\begin{document}
%
\title{Improving the identification of high-z \herschel\, sources with
  position priors and optical/NIR and FIR/mm photometric
  redshifts\thanks{\herschel\, is an ESA space observatory with
    science instruments provided by European-led Principal
    Investigator consortia and with important participation from
    NASA.}}  \titlerunning{Identification of high-z \herschel\,
  sources in the Bullet cluster field}



   \author{P.~G. P\'erez-Gonz\'alez\inst{\ref{inst1},\ref{inst2}}
     \and 
     E. Egami\inst{\ref{inst2}} 
     \and 
     M. Rex\inst{\ref{inst2}} 
     \and 
     T.~D. Rawle\inst{\ref{inst2}}  
     \and 
     J.-P. Kneib\inst{\ref{inst6}} 
     \and 
     J. Richard\inst{\ref{inst20}} 
     \and
     D. Johansson\inst{\ref{inst24}}
     \and
     B. Altieri\inst{\ref{inst8}}
     \and
     A. W. Blain\inst{\ref{inst9}}
     \and
     J. J. Bock\inst{\ref{inst9},\ref{inst10}}
     \and
     F. Boone\inst{\ref{inst11}}
     \and
     C. R. Bridge\inst{\ref{inst9}}
     \and
     S.~M. Chung\inst{\ref{inst3}} 
     \and 
     B. Cl\'ement\inst{\ref{inst6}}
     \and
     D. Clowe\inst{\ref{inst4}} 
     \and 
     F. Combes\inst{\ref{inst12}}
     \and
     J.-G. Cuby\inst{\ref{inst5}} 
     \and
     M. Dessauges-Zavadsky\inst{\ref{inst13}}
     \and 
     C. D. Dowell\inst{\ref{inst9},\ref{inst10}}
     \and
     N. Espino-Briones\inst{\ref{inst1}}
     \and
     D. Fadda\inst{\ref{inst14}}
     \and
     A. K. Fiedler\inst{\ref{inst2}} 
     \and
     A. Gonzalez\inst{\ref{inst3}} 
     \and 
     C. Horellou\inst{\ref{inst24}}
     \and
     O. Ilbert\inst{\ref{inst6}}
     \and
     R. J. Ivison\inst{\ref{inst15},\ref{inst16}}
     \and
     M. Jauzac\inst{\ref{inst6}}
     \and
     D. Lutz\inst{\ref{inst17}}
     \and
     R. Pell\'{o}\inst{\ref{inst11}}
     \and
     M. J. Pereira\inst{\ref{inst2}}
     \and
     G. H. Rieke\inst{\ref{inst2}}
     \and
     G. Rodighiero\inst{\ref{inst21}}
     \and
     D. Schaerer\inst{\ref{inst13}}
     \and
     G. P. Smith\inst{\ref{inst22}}
     \and
     I. Valtchanov\inst{\ref{inst8}}
     \and
     G. L. Walth\inst{\ref{inst2}}
     \and
     P. van der Werf\inst{\ref{inst23}}
     \and
     M. W. Werner\inst{\ref{inst10}}
     \and
     M.~Zemcov\inst{\ref{inst9},\ref{inst10}}
   }

   \institute{Departamento de Astrof\'{\i}sica, Facultad de CC. F\'{\i}sicas, Universidad Complutense de Madrid, E-28040 Madrid, Spain\label{inst1}
     \and
     Steward Observatory, The University of Arizona, 933 N Cherry Avenue, Tucson, AZ 85721, USA\label{inst2}
     \and
     Laboratoire d'Astrophysique de Marseille, CNRS - Université Aix-Marseille, 38 rue Frédéric Joliot-Curie, 13388 Marseille, France\label{inst6}
     \and
     Institute for Computational Cosmology, Department of Physics,
     Durham University, South Road, Durham DH1 3LE, UK\label{inst20}
     \and
     Onsala Space Observatory, Chalmers University of Technology, 
     SE-439 92 Onsala, Sweden\label{inst24}
     \and
     \herschel\, Science Centre, ESAC, ESA, PO Box 50727, 28080
     Madrid, Spain\label{inst8}
     \and
     California Institute of Technology, Pasadena, CA 91125,
     USA\label{inst9}
     \and
     Jet Propulsion Laboratory, Pasadena, CA 91109, USA\label{inst10}
     \and
     Laboratoire d'Astrophysique de Toulouse-Tarbes,
     Universit\'{e} de Toulouse, CNRS, 14 Av. Edouard Belin, 31400
     Toulouse, France\label{inst11}
     \and
     Department of Astronomy, University of Florida, Gainesville, FL 32611-2055, USA\label{inst3}
     \and
     Department of Physics \& Astronomy, Ohio University, Clippinger Labs 251B, Athens, OH 45701, USA\label{inst4}
     \and
     Observatoire de Paris, LERMA, 61 Av. de l'Observatoire, 75014
     Paris, France\label{inst12}
     \and
     Laboratoire d'Astrophysique de Marseille, Observatoire Astronomique de Marseille-Provence, 2 Place Le Verrier, 13248 Marseille, France\label{inst5}
     \and
     Geneva Observatory, University of Geneva, 51, Ch. des
     Maillettes, CH-1290 Versoix, Switzerland\label{inst13}
     \and
     NASA \herschel\, Science Center, California Institute of
     Technology, MS 100-22, Pasadena, CA 91125, USA\label{inst14}
     \and
     UK Astronomy Technology Centre, Science and Technology
     Facilities Council, Royal Observatory, Blackford Hill,
     Edinburgh EH9 3HJ, UK\label{inst15}
     \and
     Institute for Astronomy, University of Edinburgh, Blackford
     Hill, Edinburgh EH9 3HJ, UK\label{inst16}
     \and
     Max-Planck-Institut f\"{u}r extraterrestrische Physik,
     Postfach 1312, 85741 Garching, Germany\label{inst17}
     \and
     Department of Astronomy, University of Padova,
     Vicolo dell'Osservatorio 3, I-35122 Padova, Italy\label{inst21}
     \and
     School of Physics and Astronomy, University of Birmingham,
     Edgbaston, Birmingham, B15 2TT, UK\label{inst22}
     \and
     Sterrewacht Leiden, Leiden University, PO Box 9513, 2300 RA
     Leiden, the Netherlands\label{inst23}
   }
   
   \authorrunning{P\'erez-Gonz\'alez et al.}

   \date{Last edited: \today}

 
   \abstract{We present preliminary results about the detection of
     high redshift (U)LIRGs in the Bullet cluster field by the PACS
     and SPIRE instruments within the \herschel\, Lensing Survey (HLS)
     Program.  We describe in detail a photometric procedure designed
     to recover robust fluxes and deblend faint \herschel\, sources
     near the confusion noise. The method is based on the use of the
     positions of \spitzer/MIPS 24\mic\, sources as priors.  Our
     catalogs are able to reliably (5$\sigma$) recover galaxies with
     fluxes above 6 and 10~mJy in the PACS 100 and 160\mic\, channels,
     respectively, and 12 to 18 mJy in the SPIRE bands.  We also
     obtain spectral energy distributions covering the optical through
     the far-infrared/millimeter spectral ranges of all the
     \herschel\, detected sources, and analyze them to obtain
     independent estimations of the photometric redshift based on
     either stellar population or dust emission models. We exemplify
     the potential of the combined use of \spitzer\, position priors
     plus independent optical and IR photometric redshifts to robustly
     assign optical/NIR counterparts to the sources detected by
     \herschel\, and other (sub-)mm instruments.}


   \maketitle
%

\section{Introduction}

Based on the IRAS, ISO, and \spitzer\, missions, we know that luminous
infrared galaxies (LIRGs) experienced significant evolution in the
past 8~Gyr, roughly decreasing their typical infrared luminosity
[L(IR)] by an order of magnitude from z$\sim$1 to z$=$0
\citep[][]{1999ApJ...517..148F,2001ApJ...556..562C,
  2005ApJ...630...82P,2005ApJ...632..169L,2009arXiv0910.5649R,
  2009A&A...496...57M}.  By z$\sim$1, they dominated the star
formation activity of the Universe, being responsible for more than
50\% of the cosmic SFR density up to at least z$\sim$3, and playing an
important role in the formation of massive galaxies at z$\gtrsim$1.5,
which exhibit high star formation efficiencies
\citep{2008ApJ...675..234P,2008ApJS..175...48R, 2009ApJ...705..617D},
strong dust obscured nuclear activity
\citep{2007ApJ...670..173D,2008ApJ...687..111D}, or both. Beyond
z$\sim$1, sub-mm surveys have also detected a population of LIRGs and
(mostly) ULIRGs, identified as very active stages in the formation of
massive galaxies \citep{1997ApJ...490L...5S, 1998Natur.394..241H,
  2003Natur.422..695C, 2005ApJ...622..772C,2006MNRAS.370.1185P}.

Although very useful for measuring the obscuration of optical light by
interstellar dust (very abundant in the most extreme starbursts), the
analysis of IR data to understand the formation and evolution of
galaxies in the Universe has been classically hampered mainly by two
issues.  First, the sensitivity of IR telescopes has increased
significantly from one mission to the next (and also their angular
resolution, although at a lower rate), but there are still problems
correctly identifying all sources detected in IR surveys, most
noticeably at longer wavelengths where confusion remains severe (e.g.,
at 70/160\mic\, in the \spitzer/MIPS surveys and in the SCUBA or
LABOCA sub-mm data).  Moreover, finding counterparts and getting
reliable redshifts from spectroscopy for high redshift (U)LIRGs is
challenging since a good fraction of them are extremely faint at
optical/NIR wavelengths \citep{2002MNRAS.337....1I,
  2008MNRAS.390.1061W,2004ApJS..154..137F}. A second problem for the
study of 0$<$z$\lesssim$4 (U)LIRGs is the availability of only one
photometric data point (MIPS 24\mic) for most of them.  The fits to
dust emission models to calculate L(IR), and from this derive a SFR
(\citealt{1998ARA&A..36..189K}), need a considerable extrapolation,
and thus the SFR is subject to significant systematic uncertainties
\citep[][]{2006ApJ...640...92P, 2007ApJ...670..173D,barro10}.  In
addition, the limited FIR data already available for galaxies at
z$\gtrsim$2 indicate that they have different dust properties from
those observed for nearby (U)LIRGs
\citep{2005ApJ...622..772C,2007ApJ...668...45P,
  2008ApJ...675..262R,2009ApJ...692..556R}.





The ESA \herschel\, Space Observatory \citep{pilbratt10} alleviates
these two issues. Thanks to its sensitivity and angular resolution at
100$<$$\lambda$$<$500\mic, it enables us to robustly characterize
high-z galaxies. In addition to the populations of IR galaxies newly
discovered by \herschel, it will supply up to six photometric data
points in the IR range for 0$<$z$\lesssim$4 galaxies detected by MIPS
at 24\mic.  These data can be used to constrain the fits to the dust
emission templates.  Even more novel and relevant is the combination
of \herschel\, observations with the MIPS data (and also IRAC) to
diminish the effects of source confusion in IR surveys. The new
\herschel\, data indeed allow an easier and more reliable
identification of counterparts from one band to the adjacent one as we
move to shorter wavelengths. In addition, the \herschel\, bands
effectively fill the gap between MIPS and (sub-)mm surveys (with
SCUBA2, LABOCA, or AzTEC). By combining all these datasets, we can now
use the several IR photometric data points to estimate a photometric
redshift based on dust emission alone, and compare it with the
estimations based on UV-to-NIR data. This can be regarded as an
extension of the radio-to-IR photo-z method
\citep{1999ApJ...513L..13C,2000MNRAS.319..813D,2001PASJ...53..433R,
  2003MNRAS.338..733B,2003MNRAS.342..759A,2009ApJ...694.1517D},
benefiting from a finer sampling of the emission from dust at
different temperatures. The procedure can be used to improve the
cross-correlation with optical/NIR galaxy samples, or even to obtain
the only redshift measurement possible for very faint (high-z)
IR-bright optically-faint sources. These sources may be numerous in
the \herschel\, surveys, because of its impressive sensitivity at
wavelengths probing the peak of the dust emission, which can be
brighter than the optical emission by orders of magnitude.


In this Letter, we perform a preliminary analysis of the \herschel\,
Lensing Survey (HLS, PI Egami) data taken for the Bullet cluster to
show the utility of position priors and photometric redshifts obtained
from UV-to-NIR and IR-to-mm data to improve the identification of
\herschel\, and sub-mm sources with those detected at shorter
wavelengths. A correct identification of IR sources will be extremely
relevant to all \herschel\, cosmological surveys, but is of critical
importance for clusters, because of their crowded nature and the
possible presence of high-z lensed and distorted sources with very
faint optical counterparts that may even only be detectable in the
FIR/mm range.





\section{Data and catalogs}


The reduction of the \herschel\, SPIRE \citep{griffin10} and PACS
\citep{poglitsch10} data of the Bullet cluster ($\alpha$$\sim$06:58,
$\delta$$\sim$$-$55:57) is described in \citet{egami10} and
\citet{rex10}, jointly with optical-to-mm ancillary data used to
estimate photometric redshifts, stellar masses, and SFRs.  Here we
describe our cataloging procedure for the five \herschel\, bands: PACS
100 and 160\mic, and SPIRE 250, 350, and 500\mic.


Given the remarkable depth of the {\it Spitzer} MIPS 24~$\mu$m images,
PACS catalogs have been compiled with a position prior technique,
involving a list of MIPS sources and a PSF fitting analysis
\citep[see][]{2008ApJ...675..234P}.  To improve the reliability of the
method, we first aligned all \herschel\, images to the WCS of the MIPS
24\mic\, map using the {\it wcstools} software
\citep{1999ASPC..172..498M}, obtaining matching uncertainties smaller
than one pixel for each \herschel\, band. The MIPS 24\mic\, catalog in
the Bullet cluster field has a 5$\sigma$ detection level of
$S_{5\sigma}[24]$$=$85$\pm$30\micJy \footnote{Calculated from several
  artificial apertures formed with random sky pixels; see Appendix A
  in \citet{2008ApJ...675..234P}.}, and a density of
5.3~sources/arcmin$^2$ above that threshold (7.2~sources/arcmin$^2$
with $S[24]$$>$50\micJy).

The similar PSF sizes of the MIPS 24\mic\, and the PACS 100\mic\,
bands ($\sim$6\arcsec\, and $\sim$8\arcsec, respectively) make the
identification and extraction of sources straightforward for this
band. In addition, we also considered that some sources may be too
faint for MIPS yet still be detectable by PACS. To account for this
population, we detected sources directly in the PACS data using {\it
  Sextractor} \citep{1996A&AS..117..393B}, obtaining a sample of 43
sources above the 5$\sigma$ level
($S_{5\sigma}[100]$$=$5.5$\pm$0.7~mJy).  The cross-correlation of this
sample with the MIPS catalog (using a search radius of 3\arcsec)
revealed that all but one of the robust PACS detections are detected
at 24~\mic, and 5 of them had 2 MIPS counterparts within the search
radius. The only non-detection at 24\mic\, is, in fact, a group of
three very faint 100\mic\, sources within 10\arcsec, which are
detected as one single source by {\it Sextractor} in the PACS image,
but are individual sources at 24\mic.  The 5 multiple detections were
sources for which the slightly higher resolution and greater depth of
MIPS allowed the deblending of close IR emitters. This
cross-correlation also found that the WCS alignment between both
images has an rms smaller than 1\arcsec.


For the 160\mic\, image, it is expected that more 24\mic\, sources are
merged to form single sources. For this reason, before measuring
photometry, we removed sources from the main catalog that could not be
separated at the resolution of the PACS 160\mic\, band, i.e., sources
closer than 6\arcsec, which is half of the PSF FWHM, keeping only the
brightest source at 24\mic\, for each group of merged sources. The
master catalog combined the MIPS source list with the direct
detections in the 160\mic\, image above the 5$\sigma$ level
($S_{5\sigma}[160]$$=$10$\pm$1~mJy), all the latter being actually
detected in the MIPS data. Merged sources account for almost 20\% of
the master catalog, with 85\% (15\%) of this subsample having 2 (3)
objects within 6\arcsec.  Just considering sources directly detected
at 160\mic, the fraction of them presenting multiple identifications
in the MIPS catalog is higher than 40\%.  The WCS accuracy between the
MIPS and PACS160 images is inferior to 1.5\arcsec.

The final purged list of sources was used to fit PSFs at the given
positions using the {\it daophot} package in IRAF, allowing for one
pixel centering offsets. The results of this PSF fitting algorithm are
flux densities in a given aperture (we found the optimum values to be
5\arcsec\, and 7.5\arcsec\, for the PACS green and red channels,
respectively), to which an aperture correction must be applied. The
values we applied are 1.84$\pm$0.08 and 1.93$\pm$0.08 for 100 and
160\mic, respectively. We note that these aperture corrections are
larger than those published by the PACS Team, but we found that their
PSF matched ours (constructed with the same Bullet cluster data) for
large radii, but was brighter for similar and smaller radii than that
selected for the PSF fitting method.



\begin{table}
\caption[]{Properties of the HLS Bullet cluster catalogs.}
\label{tabledensities}
\centering
\renewcommand{\tabcolsep}{1.5mm}
\begin{tabular}{l c c c c c}
\hline
& \multicolumn{1}{c}{PACS100} & \multicolumn{1}{c}{PACS160} & \multicolumn{1}{c}{SPIRE250} & \multicolumn{1}{c}{SPIRE350} & \multicolumn{1}{c}{SPIRE500} \\
\hline
A$^{\mathrm{a}}$                   & 28.5            &   29.7          & 434             &  438      &  416                 \\
S($\lambda$)$^{\mathrm{b}}$        & 5.5 (3.0)       & 10  (6)         & 12 (7)          &  17 (10)  &  18 (11)             \\
D$^{\mathrm{c}}$                   & 2.0 (5.0)       & 2.0 (4.2)       & 1.1 (1.8)       & 0.5 (1.2) & 0.2 (0.6)            \\
M$^{\mathrm{d}}$                   & 12\% (24)       & 19\% (24)       & 48\% (24)       & 42\% (250)       & 48\% (350)      \\
\hline
\end{tabular}
\begin{list}{}{}
\item[$^{\mathrm{a}}$] Surveyed area in arcmin$^2$.
\item[$^{\mathrm{b}}$] 5$\sigma$ (3$\sigma$) flux density detection levels$^1$ in mJy.
\item[$^{\mathrm{c}}$] Source densities for each significance level in sources/arcmin$^{2}$.
\item[$^{\mathrm{d}}$] Fraction of merged sources and wavelength (in $\mu$m) used to obtain position priors.
\end{list}
\vspace{-0.9cm}
\end{table}


\begin{figure*}
   \centering
\includegraphics[angle=90,width=8.8cm]{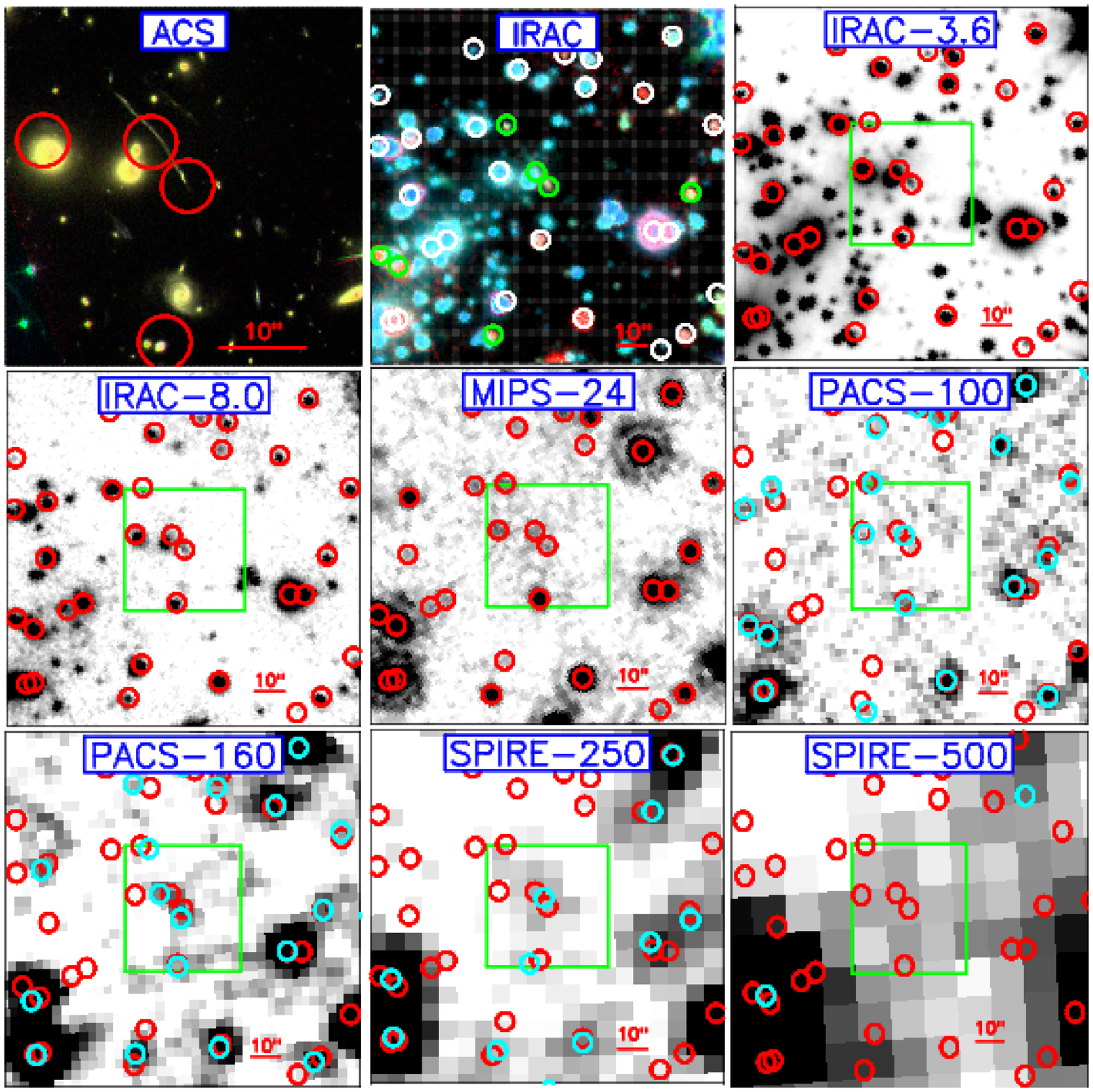}
\hspace{0.2cm}
\includegraphics[angle=90,width=8.8cm]{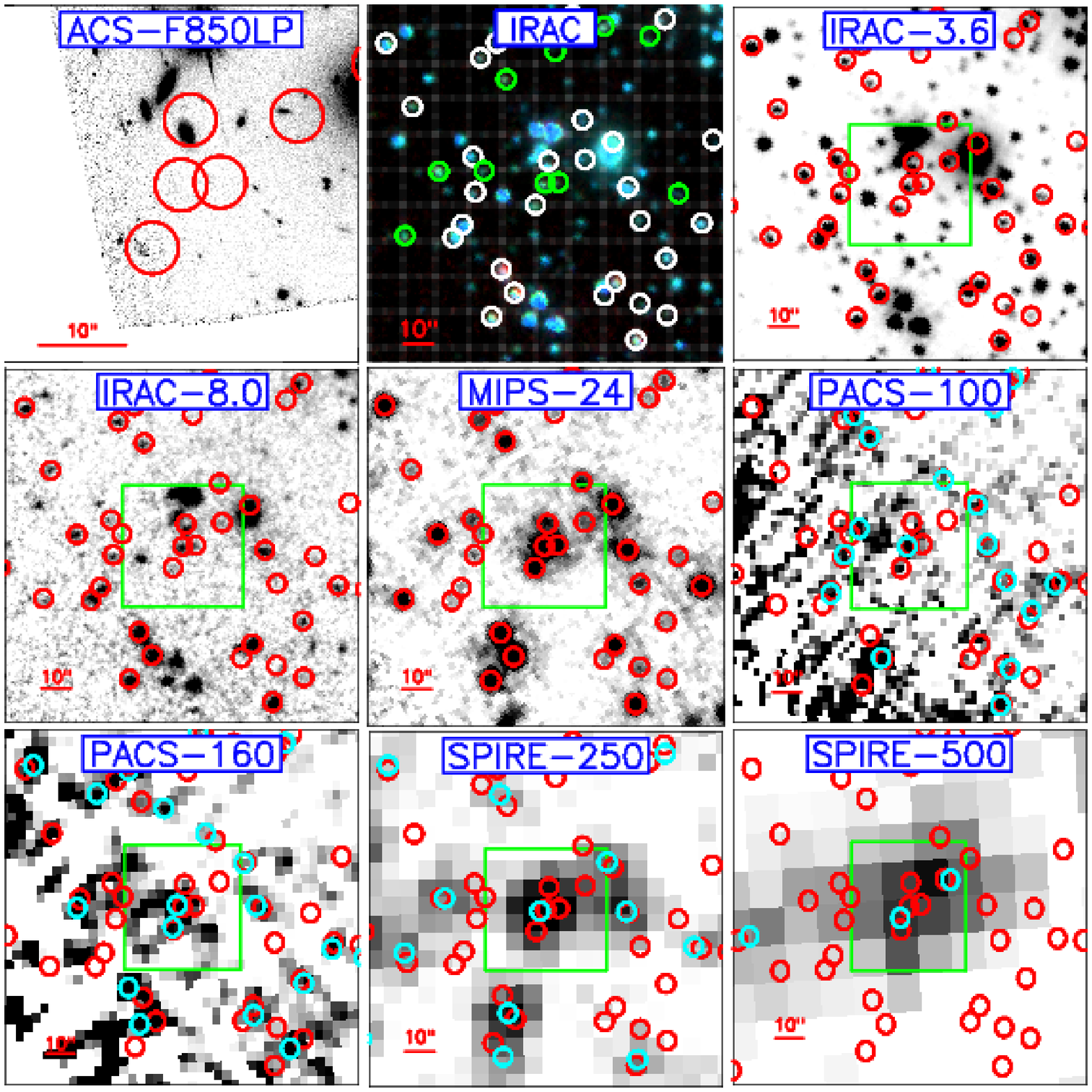}
\caption{Postage stamps of two high-z sources in the Bullet cluster
  field detected by \herschel: the southern tip of the arc
  \object{MIPS06:58:31.1$-$55:56:39.2} ({\it left}), and the LABOCA
  source \object{MIPS06:58:45.3$-$55:58:46.5} ({\it right}). The size
  of the images is 2\arcmin$\times$2\arcmin, except for the HST
  images, whose size is 40\arcsec$\times$40\arcsec (shown in the other
  stamps with a square). Circles (3\arcsec\, radius) pinpoint sources
  detected at 24\mic. Cyan circles show the actual $>$5$\sigma$
  detections in each \herschel\, band. Green circles in the IRAC RGB
  images mark z$>$2 galaxies in each field, including the three lensed
  sources in \citet[][to the SE or the arc]{2009ApJ...691..525G} and a
  LABOCA source from \citet[][to the W of the
  arc]{2010arXiv1003.0827J}.  }
  \label{postages}
\end{figure*}

The SPIRE catalogs were also compiled with a method based on priors
and PSF fitting. The photometry was carried out in circular apertures
(of radii 12\arcsec, 18\arcsec, and 22\arcsec\, for 250, 350, and
500\mic, respectively), applying a calibration based on the beam sizes
(assuming they are Gaussians). For the 250\mic\, channel, we combined
the list of sources detected by MIPS at 24\mic\, and the direct
detection in the SPIRE data using {\it Sextractor} and a 5$\sigma$
threshold ($S_{5\sigma}[250]$$=$12$\pm$2~mJy). Other objects within
9\arcsec\, (half of the PSF FWHM) of a given candidate source were
purged, only the position of the brightest galaxy at 24\mic\, being
kept.  For the region covered simultaneously by MIPS and the SPIRE
250\mic\, channel, only 2 SPIRE sources out of 261 were not detected
by MIPS, but both of them turned to be F(24)$\sim$60\micJy\,
detections at the 3$\sigma$ level.  Around half of the SPIRE sources
have 2 or more MIPS counterparts within 9\arcsec. The WCS alignment
between the MIPS and SPIRE250 images has an rms of $\sim$4\arcsec.


The catalog for the SPIRE 350\mic\, channel was constructed from the
list of detections at 250\mic. Sources within 14\arcsec\, were merged
into the same object (about 40\% of the total catalog, virtually all
of them being a doublet), only the position of the brightest one at
250\mic\, being kept.  A direct detection in the 350\mic\, image
($S_{5\sigma}[350]$$=$17$\pm$3~mJy) found that 5\% of sources are not
directly linked to a 250\mic\, emitter, but virtually all of these are
a combination of several 250\mic\, sources that causes a displaced
350\mic\, centroid.  Only one isolated high-confidence 350\mic\,
source was not detected in the 250\mic\, catalog, but a reliable flux
could be recovered in the blue channel once the 350\mic\, detection
revealed its position.  The WCS alignment between the MIPS and
SPIRE350 images has an rms of $\sim$6\arcsec.

For the 500\mic\, channel ($S_{5\sigma}[500]$$=$18$\pm$4~mJy), we
started from the 350\mic\, catalog, and we did not find any object
without an identification in the bluer bands using a 18\arcsec\,
search radius (approximately half of the PSF FWHM). Around 50\% of the
SPIRE 500\mic\, sources were doublets in the 350\mic\, band.  The WCS
alignment between the MIPS and SPIRE500 images has an rms of
$\sim$9\arcsec.

The prior-based photometric technique was compared with the direct
blind detection in the \herschel\, images.  In PACS, a direct
detection down to 5$\sigma$ included more than 40\% of sources with no
counterpart in the MIPS map, implying that these are probably spurious
detections. The fluxes of sources extracted with both the blind and
prior-based detections are virtually identical.  In SPIRE, the
prior-based procedure is able to recover fluxes for 30\% more sources
than the direct detection.  For the sources in common, the average
difference in the photometry is 2\% (the blind method provides
brighter fluxes), with a 6\% scatter.






Table~\ref{tabledensities} indicates the surveyed areas, detection
thresholds, the corresponding source densities, and fraction of merged
sources for our catalogs of the PACS and SPIRE HLS data. An important
(and potentially dangerous) step in our photometric procedure based on
position priors is the blending of nearby non-resolved sources, which
provides only the position of the brightest galaxy in the parent
catalog to the PSF fitting algorithm. We note that the MIPS 24\mic\,
flux is dominated by the emission of warm dust and/or PAHs while the
SPIRE bands are dominated by the emission of cold dust.  Thus, it is
probable that bright MIR emitters are not the dominant sources at FIR
wavelengths (see Figure~\ref{postages}). In the following section, we
illustrate a method to help in the robust identification and/or
deblending of the \herschel\, sources.


\section{Identification of high-z
  \herschel\, sources}


Figure~\ref{postages} shows thumbnails in several bands of two
interesting sources detected by \herschel\, in the Bullet cluster
field: the southern tip of the arc at z$=$3.24 described in
\citet{2001A&A...379...96M}, and a sub-mm galaxy detected by LABOCA
\citep[source \#10 in][]{2010arXiv1003.0827J}.  With these two
examples, we demonstrate that the use of IRAC and MIPS data helps to
correctly identify the \herschel\, sources with optical/NIR
counterparts. In the case of the arc, the photometric method based on
the use of the MIPS 24\mic\, catalog as a prior helped in the
detection and measurement processes for the PACS 160\mic\, and SPIRE
bands, where the source is very faint and almost indistinguishable
from the background.  In the PACS red band, our method allows a
deblending of the source in the two knots detected by MIPS and linked
to the arc.  In the case of the LABOCA source, although it is barely
detected in the optical, the position of the source can be followed as
we move to redder wavelengths and larger PSFs, allowing a more
reliable identification in spite of the presence of a close neighbor
to the west. The brightest source at 24\mic\, is indeed already
brighter than its companion at IRAC wavelengths (most noticeably, at
8.0\mic), and this nearby source disappears in PACS.  This illustrates
the benefits of the use of \spitzer\, data as priors to obtain the
most probable identifications of \herschel\, and (sub-)mm emitters at
shorter wavelengths.

\begin{figure}
   \centering
\includegraphics[angle=-90,width=7.7cm]{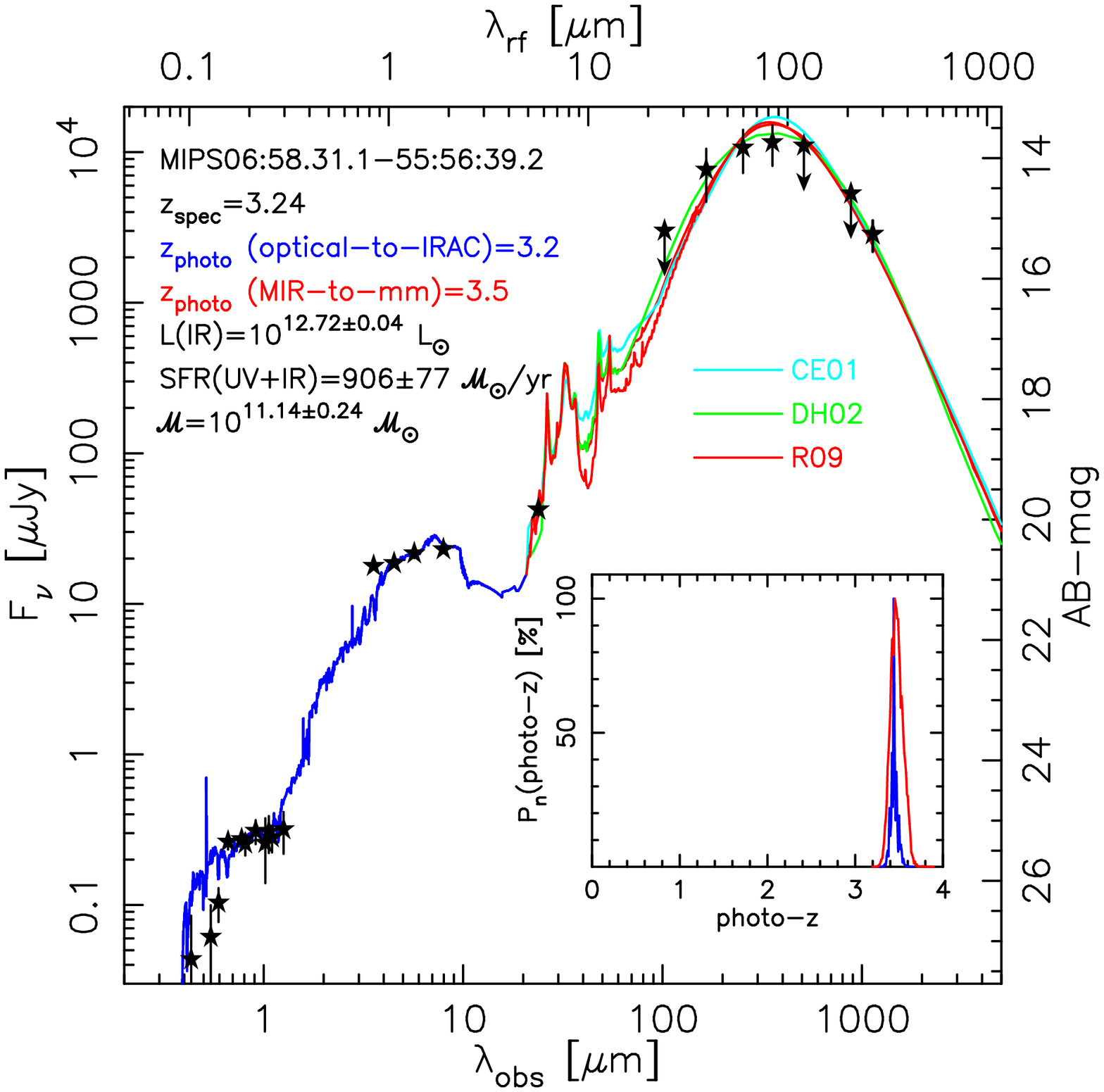}
\includegraphics[angle=-90,width=7.7cm]{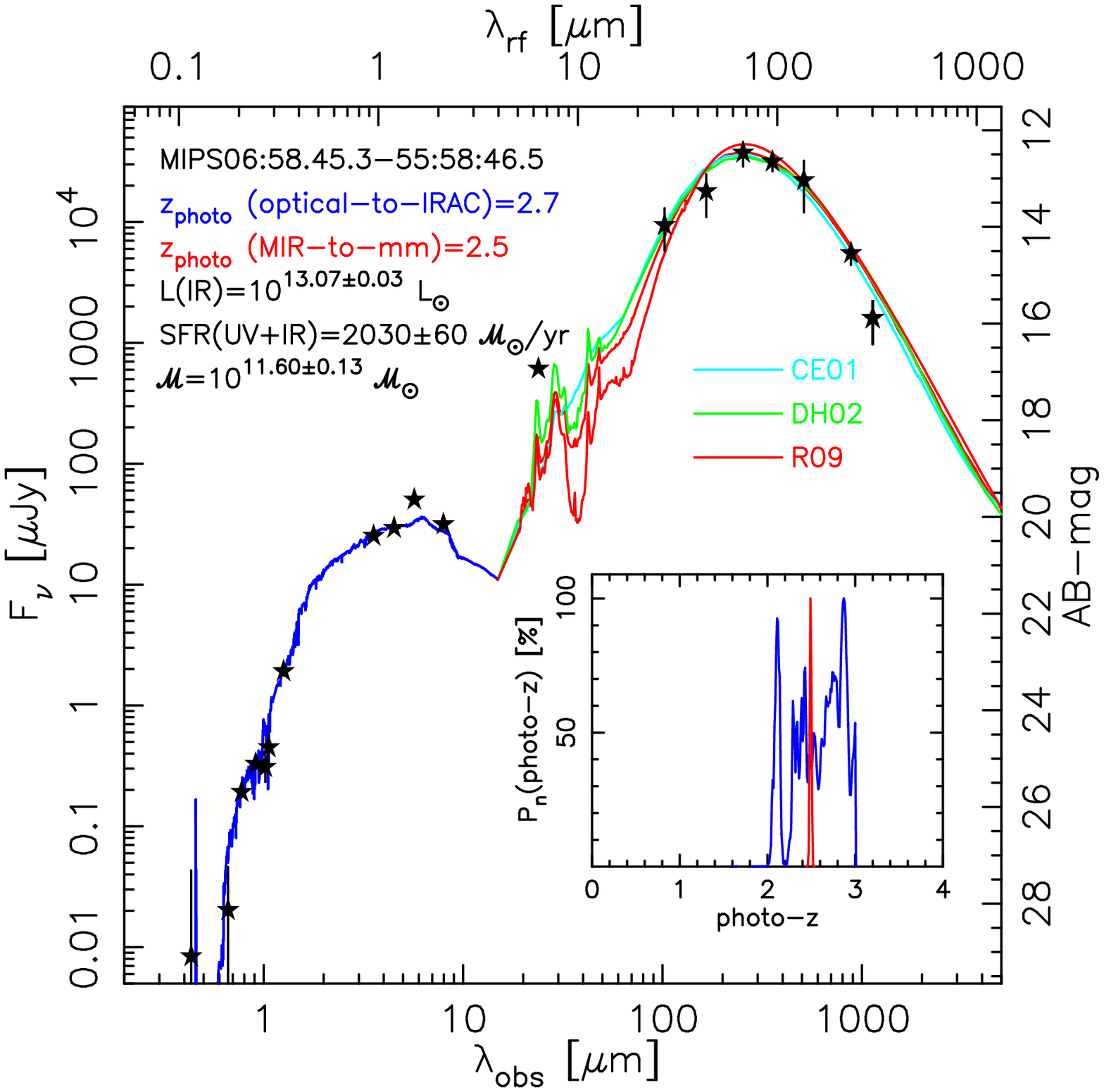}
\caption{SEDs of the z$=$3.24 LIRG
  \object{MIPS06:58:31.1$-$55:56:39.2} (magnification factor $\sim$21)
  and the z$\sim$2.5 HyLIRG \object{MIPS06:58:45.3$-$55:58:46.5},
  detected in the MIR/FIR by \spitzer\, and \herschel, and the
  (sub-)mm by LABOCA and AzTEC. The blue line shows the best-fit
  stellar population model for the photometry up to 8\mic, which
  provides an estimate of the photometric redshift and the stellar
  mass. The cyan, green, and red lines show the best-fit dust emission
  model for the MIR-to-mm photometry (from
  \citealt{2001ApJ...556..562C}, \citealt{2002ApJ...576..159D}, and
  \citealt{2009ApJ...692..556R}), which gives an independent
  photometric redshift estimation and a SFR value (given in the plots
  with no correction for magnification factor).  The inset plots
  depict the photometric redshift normalized probability distributions
  for the optical/NIR and the MIR/mm estimations.}
\vspace{-0.4cm}
\label{sed1}
\end{figure}

Figure~\ref{sed1} shows the optical-to-mm SEDs of the two examples of
high-z galaxies. Photometric redshifts for all 24\mic\, sources in the
Bullet cluster field were estimated from the UV-to-MIR data (including
IRAC fluxes) using the technique described in
\citet{2005ApJ...630...82P, 2008ApJ...675..234P}. Briefly, the SEDs
compiled from aperture-matched photometry are fitted with stellar
population models and AGN templates from \citet{2007ApJ...663...81P}.
A minimization algorithm is used to search for the optimal redshifted
template fitting the data, and the most probable photometric redshift
is calculated by integrating the redshift probability function (see
also \citealt{barro10}).  Photometric redshifts were also estimated
from the IR data alone by comparing the \spitzer, \herschel\, and
(sub-)mm fluxes with dust emission templates from
\citet{2001ApJ...556..562C}, \citet{2002ApJ...576..159D}, and
\citet{2009ApJ...692..556R}. Using these two fits, we estimated
relevant parameters such as the stellar mass and the SFR. The
\citet{2009ApJ...692..556R} template most closely fitting the data for
the arc corresponds to a local L(IR)$=$10$^{11.5}$~L$_\odot$ galaxy.
This value is consistent with the average for the arc based on the
three template libraries, L(IR)$=$10$^{11.40\pm0.04}$~L$_\odot$, once
a magnification factor of $\sim$21 is applied
\citep{2006ApJ...652..937B}. The LABOCA source qualifies as a
hyper-LIRG based on the best fits to all template libraries (if it is
not lensed), and the best-fit template from Rieke et al. is that of a
local galaxy with 10$^{12.75}$~L$_\odot$, although our source seems to
have more prominent PAH emission.




Figure~\ref{sed1} illustrates the power of using independent estimates
of the photometric redshift based on UV-to-NIR data and IR data alone
to validate the individual values (see the photo-z probability
distributions). The procedure is also useful for achieving more
reliable identifications of the IR emitters. For example, for the
LABOCA source at z$\sim$2.5, we calculated photometric redshifts based
on UV-to-NIR data for the three close neighbors detected by MIPS
within $\sim$10\arcsec\, of the central source. The galaxy to the
north lies at z$\sim$0.3 (a cluster member), that to the south at
z$\sim$1.9, and the closest companion has a photo-z compatible with
the central source (z$\sim$2.3$\pm$0.3), thus implying that the two
sources are a possible interacting pair.

The general validity of our method for the whole sample of \herschel\,
sources will be tested thoroughly in forthcoming papers by evaluating
the quality of our photo-z's. From a preliminary comparison of the
optical and FIR photo-z's for sources detected by MIPS, PACS, SPIRE,
and a (sub)-mm instrument, it is encouraging that we measure an
average $\Delta(z)/(1+z)$$=$$\eta$$=$0.09 with 10\% of catastrophic
outliers ($\eta$$>$0.2), comparable to the typical goodness of
IR-based photo-z's \citep{2003MNRAS.342..759A}.




\begin{acknowledgements}
  This work is based in part on observations made with {\it Spitzer},
  operated by JPL/Caltech.  We thank the AzTEC Team for letting us use
  their data. PGP-G acknowledges support from grants AYA 2006--02358,
  AYA 2006--15698--C02--02, and CSD2006-00070, and the Ram\'on y Cajal
  Program, all financed by the Spanish Government and/or the European
  Union.
\end{acknowledgements}




\bibliographystyle{aa}
\bibliography{referencias}

\begin{thebibliography}{44}
\expandafter\ifx\csname natexlab\endcsname\relax\def\natexlab#1{#1}\fi

\bibitem[{{Aretxaga} {et~al.}(2003){Aretxaga}, {Hughes}, {Chapin},
  {Gazta{\~n}aga}, {Dunlop}, \& {Ivison}}]{2003MNRAS.342..759A}
{Aretxaga}, I., {Hughes}, D.~H., {Chapin}, E.~L., {et~al.} 2003, \mnras, 342,
  759

\bibitem[{{Barro et al.}(2010)}]{barro10}
{Barro et al.} 2010, ApJ (submitted)

\bibitem[{{Bertin} \& {Arnouts}(1996)}]{1996A&AS..117..393B}
{Bertin}, E. \& {Arnouts}, S. 1996, \aaps, 117, 393

\bibitem[{{Blain} {et~al.}(2003){Blain}, {Barnard}, \&
  {Chapman}}]{2003MNRAS.338..733B}
{Blain}, A.~W., {Barnard}, V.~E., \& {Chapman}, S.~C. 2003, \mnras, 338, 733

\bibitem[{{Brada{\v c}} {et~al.}(2006){Brada{\v c}}, {Clowe}, {Gonzalez},
  {Marshall}, {Forman}, {Jones}, {Markevitch}, {Randall}, {Schrabback}, \&
  {Zaritsky}}]{2006ApJ...652..937B}
{Brada{\v c}}, M., {Clowe}, D., {Gonzalez}, A.~H., {et~al.} 2006, \apj, 652,
  937

\bibitem[{{Carilli} \& {Yun}(1999)}]{1999ApJ...513L..13C}
{Carilli}, C.~L. \& {Yun}, M.~S. 1999, \apjl, 513, L13

\bibitem[{{Chapman} {et~al.}(2003){Chapman}, {Blain}, {Ivison}, \&
  {Smail}}]{2003Natur.422..695C}
{Chapman}, S.~C., et~al., I.~R. 2003,
  \nat, 422, 695

\bibitem[{{Chapman} {et~al.}(2005){Chapman}, {Blain}, {Smail}, \&
  {Ivison}}]{2005ApJ...622..772C}
{Chapman}, S.~C., {Blain}, A.~W., {Smail}, I., \& {Ivison}, R.~J. 2005, \apj,
  622, 772

\bibitem[{{Chary} \& {Elbaz}(2001)}]{2001ApJ...556..562C}
{Chary}, R. \& {Elbaz}, D. 2001, \apj, 556, 562

\bibitem[{{Daddi} {et~al.}(2007){Daddi}, {Alexander}, {Dickinson}, {Gilli},
  {Renzini}, {Elbaz}, {Cimatti}, {Chary}, {Frayer}, {Bauer}, {Brandt},
  {Giavalisco}, {Grogin}, {Huynh}, {Kurk}, {Mignoli}, {Morrison}, {Pope}, \&
  {Ravindranath}}]{2007ApJ...670..173D}
{Daddi}, E., {Alexander}, D.~M., {Dickinson}, M., {et~al.} 2007, \apj, 670, 173

\bibitem[{{Daddi} {et~al.}(2009){Daddi}, {Dannerbauer}, {Stern}, {Dickinson},
  {Morrison}, {Elbaz}, {Giavalisco}, {Mancini}, {Pope}, \&
  {Spinrad}}]{2009ApJ...694.1517D}
{Daddi}, E., {Dannerbauer}, H., {Stern}, D., {et~al.} 2009, \apj, 694, 1517

\bibitem[{{Dale} \& {Helou}(2002)}]{2002ApJ...576..159D}
{Dale}, D.~A. \& {Helou}, G. 2002, \apj, 576, 159

\bibitem[{{Damen} {et~al.}(2009){Damen}, {F{\"o}rster Schreiber}, {Franx},
  {Labb{\'e}}, {Toft}, {van Dokkum}, \& {Wuyts}}]{2009ApJ...705..617D}
{Damen}, M., {F{\"o}rster Schreiber}, N.~M., {Franx}, M., {et~al.} 2009, \apj,
  705, 617

\bibitem[{{Donley} {et~al.}(2008){Donley}, {Rieke}, {P{\'e}rez-Gonz{\'a}lez},
  \& {Barro}}]{2008ApJ...687..111D}
{Donley}, J.~L., et~al.,
  G. 2008, \apj, 687, 111

\bibitem[{{Dunne} {et~al.}(2000){Dunne}, {Clements}, \&
  {Eales}}]{2000MNRAS.319..813D}
{Dunne}, L., {Clements}, D.~L., \& {Eales}, S.~A. 2000, \mnras, 319, 813

\bibitem[{{Egami et al.}(2010)}]{egami10}
{Egami et al.} 2010, \aap (this volume)

\bibitem[{{Flores} {et~al.}(1999){Flores}, {Hammer}, {Thuan}, {C{\' e}sarsky},
  {Desert}, {Omont}, {Lilly}, {Eales}, {Crampton}, \& {Le F{\`
  e}vre}}]{1999ApJ...517..148F}
{Flores}, H., {Hammer}, F., {Thuan}, T.~X., {et~al.} 1999, \apj, 517, 148

\bibitem[{{Frayer} {et~al.}(2004){Frayer}, {Chapman}, {Yan}, {Armus}, {Helou},
  {Fadda}, {Morganti}, {Garrett}, {Appleton}, {Choi}, {Fang}, {Heinrichsen},
  {Im}, {Lacy}, {Marleau}, {Masci}, {Shupe}, {Soifer}, {Squires},
  {Storrie-Lombardi}, {Surace}, {Teplitz}, \& {Wilson}}]{2004ApJS..154..137F}
{Frayer}, D.~T., {Chapman}, S.~C., {Yan}, L., {et~al.} 2004, \apjs, 154, 137

\bibitem[{{Gonzalez} {et~al.}(2009){Gonzalez}, {Clowe}, {Brada{\v c}},
  {Zaritsky}, {Jones}, \& {Markevitch}}]{2009ApJ...691..525G}
{Gonzalez}, A.~H., {Clowe}, D., {Brada{\v c}}, M., {et~al.} 2009, \apj, 691,
  525

\bibitem[{{Griffin et al.}(2010)}]{griffin10}
{Griffin et al.} 2010, \aap (this volume)

\bibitem[{{Hughes} {et~al.}(1998){Hughes}, {Serjeant}, {Dunlop},
  {Rowan-Robinson}, {Blain}, {Mann}, {Ivison}, {Peacock}, {Efstathiou}, {Gear},
  {Oliver}, {Lawrence}, {Longair}, {Goldschmidt}, \&
  {Jenness}}]{1998Natur.394..241H}
{Hughes}, D.~H., {Serjeant}, S., {Dunlop}, J., {et~al.} 1998, \nat, 394, 241

\bibitem[{{Ivison} {et~al.}(2002){Ivison}, {Greve}, {Smail}, {Dunlop}, {Roche},
  {Scott}, {Page}, {Stevens}, {Almaini}, {Blain}, {Willott}, {Fox}, {Gilbank},
  {Serjeant}, \& {Hughes}}]{2002MNRAS.337....1I}
{Ivison}, R.~J., {Greve}, T.~R., {Smail}, I., {et~al.} 2002, \mnras, 337, 1

\bibitem[{{Johansson} {et~al.}(2010){Johansson}, {Horellou}, {Sommer}, {Basu},
  {Bertoldi}, {Birkinshaw}, {Lancaster}, {Lopez-Cruz}, \&
  {Quintana}}]{2010arXiv1003.0827J}
{Johansson}, D., {Horellou}, C., {Sommer}, M.~W., {et~al.} 2010, \aap\, (accepted), astro-ph/1003.0827

\bibitem[{{Kennicutt}(1998)}]{1998ARA&A..36..189K}
{Kennicutt}, R.~C. 1998, \araa, 36, 189

\bibitem[{{Le Floc'h} {et~al.}(2005){Le Floc'h}, {Papovich}, {Dole}, {Bell},
  {Lagache}, {Rieke}, {Egami}, {P{\'e}rez-Gonz{\'a}lez}, {Alonso-Herrero},
  {Rieke}, {Blaylock}, {Engelbracht}, {Gordon}, {Hines}, {Misselt}, {Morrison},
  \& {Mould}}]{2005ApJ...632..169L}
{Le Floc'h}, E., {Papovich}, C., {Dole}, H., {et~al.} 2005, \apj, 632, 169

\bibitem[{{Magnelli} {et~al.}(2009){Magnelli}, {Elbaz}, {Chary}, {Dickinson},
  {Le Borgne}, {Frayer}, \& {Willmer}}]{2009A&A...496...57M}
{Magnelli}, B., {Elbaz}, D., {Chary}, R.~R., {et~al.} 2009, \aap, 496, 57

\bibitem[{{Mehlert} {et~al.}(2001){Mehlert}, {Seitz}, {Saglia}, {Appenzeller},
  {Bender}, {Fricke}, {Hoffmann}, {Hopp}, {Kudritzki}, \&
  {Pauldrach}}]{2001A&A...379...96M}
{Mehlert}, D., {Seitz}, S., {Saglia}, R.~P., {et~al.} 2001, \aap, 379, 96

\bibitem[{{Mink}(1999)}]{1999ASPC..172..498M}
{Mink}, D.~J. 1999, in Astronomical Society of the Pacific Conference Series,
  Vol. 172, Astronomical Data Analysis Software and Systems VIII, ed.
  {D.~M.~Mehringer, R.~L.~Plante, \& D.~A.~Roberts}, 498--+

\bibitem[{{P{\' e}rez-Gonz{\' a}lez} {et~al.}(2005){P{\' e}rez-Gonz{\' a}lez},
  {Rieke}, {Egami}, {Alonso-Herrero}, {Dole}, {Papovich}, {Blaylock}, {Jones},
  {Rieke}, {Rigby}, {Barmby}, {Fazio}, {Huang}, \&
  {Martin}}]{2005ApJ...630...82P}
{P{\' e}rez-Gonz{\' a}lez}, P.~G., {Rieke}, G.~H., {Egami}, E., {et~al.} 2005,
  \apj, 630, 82

\bibitem[{{P{\' e}rez-Gonz{\' a}lez} {et~al.}(2008){P{\' e}rez-Gonz{\' a}lez},
  {Rieke}, {Villar}, {Barro}, {Blaylock}, {Egami}, {Gallego}, {Gil de Paz},
  {Pascual}, {Zamorano}, \& {Donley}}]{2008ApJ...675..234P}
{P{\' e}rez-Gonz{\' a}lez}, P.~G., {Rieke}, G.~H., {Villar}, V., {et~al.} 2008,
  \apj, 675, 234

\bibitem[{{Papovich} {et~al.}(2006){Papovich}, {Moustakas}, {Dickinson}, {Le
  Floc'h}, {Rieke}, {Daddi}, {Alexander}, {Bauer}, {Brandt}, {Dahlen}, {Egami},
  {Eisenhardt}, {Elbaz}, {Ferguson}, {Giavalisco}, {Lucas}, {Mobasher},
  {P{\'e}rez-Gonz{\'a}lez}, {Stutz}, {Rieke}, \& {Yan}}]{2006ApJ...640...92P}
{Papovich}, C., {Moustakas}, L.~A., {Dickinson}, M., {et~al.} 2006, \apj, 640,
  92

\bibitem[{{Papovich} {et~al.}(2007){Papovich}, {Rudnick}, {Le Floc'h}, {van
  Dokkum}, {Rieke}, {Taylor}, {Armus}, {Gawiser}, {Huang}, {Marcillac}, \&
  {Franx}}]{2007ApJ...668...45P}
{Papovich}, C., {Rudnick}, G., {Le Floc'h}, E., {et~al.} 2007, \apj, 668, 45

\bibitem[{{Pilbratt et al.}(2010)}]{pilbratt10}
{Pilbratt et al.} 2010, \aap (this volume)

\bibitem[{{Poglitsch et al.}(2010)}]{poglitsch10}
{Poglitsch et al.} 2010, \aap (this volume)

\bibitem[{{Polletta} {et~al.}(2007){Polletta}, {Tajer}, {Maraschi},
  {Trinchieri}, {Lonsdale}, {Chiappetti}, {Andreon}, {Pierre}, {Le F{\`e}vre},
  {Zamorani}, {Maccagni}, {Garcet}, {Surdej}, {Franceschini}, {Alloin},
  {Shupe}, {Surace}, {Fang}, {Rowan-Robinson}, {Smith}, \&
  {Tresse}}]{2007ApJ...663...81P}
{Polletta}, M., {Tajer}, M., {Maraschi}, L., {et~al.} 2007, \apj, 663, 81

\bibitem[{{Pope} {et~al.}(2006){Pope}, {Scott}, {Dickinson}, {Chary},
  {Morrison}, {Borys}, {Sajina}, {Alexander}, {Daddi}, {Frayer}, {MacDonald},
  \& {Stern}}]{2006MNRAS.370.1185P}
{Pope}, A., {Scott}, D., {Dickinson}, M., {et~al.} 2006, \mnras, 370, 1185

\bibitem[{{Reddy} {et~al.}(2008){Reddy}, {Steidel}, {Pettini}, {Adelberger},
  {Shapley}, {Erb}, \& {Dickinson}}]{2008ApJS..175...48R}
{Reddy}, N.~A., {Steidel}, C.~C., {Pettini}, M., {et~al.} 2008, \apjs, 175, 48

\bibitem[{{Rengarajan} \& {Takeuchi}(2001)}]{2001PASJ...53..433R}
{Rengarajan}, T.~N. \& {Takeuchi}, T.~T. 2001, \pasj, 53, 433

\bibitem[{{Rex et al.}(2010)}]{rex10}
{Rex et al.} 2010, \aap (this volume)

\bibitem[{{Rieke} {et~al.}(2009){Rieke}, {Alonso-Herrero}, {Weiner},
  {P{\'e}rez-Gonz{\'a}lez}, {Blaylock}, {Donley}, \&
  {Marcillac}}]{2009ApJ...692..556R}
{Rieke}, G.~H., {Alonso-Herrero}, A., {Weiner}, B.~J., {et~al.} 2009, \apj,
  692, 556

\bibitem[{{Rigby} {et~al.}(2008){Rigby}, {Marcillac}, {Egami}, {Rieke},
  {Richard}, {Kneib}, {Fadda}, {Willmer}, {Borys}, {van der Werf},
  {P{\'e}rez-Gonz{\'a}lez}, {Knudsen}, \& {Papovich}}]{2008ApJ...675..262R}
{Rigby}, J.~R., {Marcillac}, D., {Egami}, E., {et~al.} 2008, \apj, 675, 262

\bibitem[{{Rodighiero} {et~al.}(2010){Rodighiero}, {Vaccari}, {Franceschini},
  {Tresse}, {Le Fevre}, {Le Brun}, {Mancini}, {Matute}, {Cimatti}, {Marchetti},
  {Ilbert}, {Arnouts}, {Bolzonella}, {Zucca}, {Bardelli}, {Lonsdale}, {Shupe},
  {Surace}, {Rowan-Robinson}, {Garilli}, {Zamorani}, {Pozzetti}, {Bondi}, {de
  la Torre}, {Vergani}, {Santini}, {Grazian}, \&
  {.~Fontana}}]{2009arXiv0910.5649R}
{Rodighiero}, G., {Vaccari}, M., {Franceschini}, A., {et~al.} 2010, \aap\, (accepted), astro-ph/0910.5649

\bibitem[{{Smail} {et~al.}(1997){Smail}, {Ivison}, \&
  {Blain}}]{1997ApJ...490L...5S}
{Smail}, I., {Ivison}, R.~J., \& {Blain}, A.~W. 1997, \apjl, 490, L5

\bibitem[{{Wilson} {et~al.}(2008){Wilson}, {Hughes}, {Aretxaga}, {Ezawa},
  {Austermann}, {Doyle}, {Ferrusca}, {Hern{\'a}ndez-Curiel}, {Kawabe},
  {Kitayama}, {Kohno}, {Kuboi}, {Matsuo}, {Mauskopf}, {Murakoshi},
  {Monta{\~n}a}, {Natarajan}, {Oshima}, {Ota}, {Perera}, {Rand}, {Scott},
  {Tanaka}, {Tsuboi}, {Williams}, {Yamaguchi}, \& {Yun}}]{2008MNRAS.390.1061W}
{Wilson}, G.~W., {Hughes}, D.~H., {Aretxaga}, I., {et~al.} 2008, \mnras, 390,
  1061

\end{thebibliography}

\end{document}